\documentstyle[prd,aps,floats]{revtex}
\begin{document}
\draft
\title{Nucleation Rate of Hadron Bubbles in Baryon-Free Quark-Gluon
Plasma}
\author{Franco Ruggeri and William Friedman}
\address{Physics Department, University of Wisconsin, Madison,
Wisconsin}
\twocolumn[
\date{\today}
\maketitle
\widetext
\begin{abstract}
\begin{center}
\parbox{14cm}{
We evaluate the factor $\kappa$ appearing in Langer's expression for
the nucleation rate extended to the case of hadron bubbles forming
in
zero baryon number cooled quark-gluon plasma. We consider both the
absence and presence of viscosity and show that viscous effects
introduce only small changes in the value of $\kappa$.}
\end{center}
\end{abstract}
\pacs{\hspace{1.9cm} PACS numbers: 24.85.+p, 12.38.Mh, 25.75.+r,
64.60.Qb}
]
\narrowtext

\section{Introduction}

    There has been a great deal of recent activity related to the
study of the quark-hadron
phase transition. For the transition from quark matter to hadron
matter there is interest in the rate of formation of critical size
hadron bubbles, specifically bubbles of pion gas, in a supercooled
vapor [1,2].
The nucleation rate is an essential ingredient of the complete
treatment of a  phase transition, as critical size bubbles must
first be formed in the quark vapor before the vapor can be converted
into hadron material.
 Hadron bubbles initially form due to fluctuations of the
energy density in the vapor. Those with radii, R, smaller than a
critical size $R_o$ collapse, while those of the critical size begin
to grow exponentially. Langer and Turski [3] have shown that the
nucleation rate, $I$, can be written as follows:
\begin{equation}
I=\frac{\kappa}{2\pi} \Omega_o {\rm exp}(-\Delta F/T).
\end{equation}
Here $\kappa$ is related to the growth rate of the radius, $R$, of
the
bubble near the critical radius, $R_o$, by $dR/dt \approx \kappa
(R-R_o)$. The factor $\Omega_o$ is a statistical prefactor, and
$\Delta F$ the
difference in free energy of systems with, and without, a critical
size drop present. This paper is concerned primarily with discussing
the rate  $\kappa$.

Csernai and Kapusta [1] have applied the Langer-Turski nucleation
rate, (1), to the formation of hadron bubbles in a quark vapor. They
consider both
the hadron and quark materials as substances with zero
baryon number and treat them using relativistic formalism. One
principal result of [1] is the suggestion that for baryon-free matter
 $\kappa$ is
proportional to
the  viscosity
coefficients of the quark vapor, and thus when these coefficients
vanish a hadron
 bubble does not form.

We reexamine the relativistic evaluation of $\kappa$ in this paper,
and show
that viscosity is not necessary for the growth of hadron bubbles, and
that its effects are of higher order than the energy flow terms in
the
growth process. The difference between our results and those of
Ref. [1] seems to arise from the conceptual
distinction between heat conduction and energy
flow.  The former is, indeed, not
defined in baryon free systems i.e. systems with a chemical potential
of zero as pointed out in [1] and other sources. Energy flow,
however,
is defined. A difference between our results and Ref. [1] also
 arises from the technical difference in the treatment of the
pressure gradients.

The paper is organized as follows. In Section 2., we review the
treatment of
energy flow, viscosity and thermal conduction in the baryon, and
baryon-
free, cases of relativistic hydrodynamics. We consider the
distinction
between heat conduction and energy flow and stress that the latter
can
occur without viscosity in the baryon-free case. In
Section 3., we calculate the rate
 $\kappa$ in the absence of
viscosity and baryon number by solving  relativistic energy and
momentum
 transport equations. The energy
and momentum relativistic equations are mathematically of the same
form as the mass conservation and momentum nonrelativistic transport
equations for a nonviscous vapor with perfect heat conduction. We
compare the relativistic result with
 the solution for
$\kappa$ in such a nonrelativistic case as obtained in [3],
and show that the forms are similar. The method
of solution of
  the transport equations
used in this paper differs from that used in Ref. [3].
We compare the two methods, both of which can be used in
either
 the relativistic and nonrelativistic cases. In Section 4, we
 reexamine the calculation of the rate $\kappa$ in [1], focusing
on the relationship between the free energy and
the gradient of pressure, which
is used in the
momentum transport
equation.
In particular, we suggest that a source for the difference between
the
conclusion of this work and those of Ref. [1], is linked
to a difference in treatments of the pressure gradients.
Finally, in Section 5, we examine the relativistic transport
equations in the presence of viscosity and suggest that viscous
effects
are of high order in the case of small viscosity coefficients.

\section{Energy versus Heat Flow in Baryon and Baryon-Free
Relativistic Hydrodynamics}

 We first review the transport relationships for a system
with nonzero baryon number. This is a system
for which the
 baryon chemical potential is also nonzero. Under this condition
there
is a net number of either particles or antiparticles present
. Thus one can define a number density, n, and
obtain the conservation of net particle (or antiparticle) number in
the
 absence of
dissipative effects:
\begin{equation}
\label{nflow}
\frac{\partial( nu^i)}{\partial x^i} = 0.
\end{equation}
Here $u^i$ represents the 4-velocity (divided by the velocity of
light) of the net particles.
The energy density and flow are given by the stress tensor (also in
the
absence of dissipative effects):
\begin{equation}
\label{T}
T^{ij} = pg^{ij} + (e+p)u^iu^j
\end{equation}
which provides energy and momentum conservation through the
relationship:
\begin{equation}
\label{flo}
\frac{\partial T^{ij}}{\partial x^i}=0.
\end{equation}
Here $g^{ik}$ is the metric, $e$, the energy density, $p$, pressure,
 and $u_0 = 1/ \sqrt
(1-v^2/c^2)=\gamma$. The combination $e+p$ is the enthalpy
density, $w$. Both the energy flow and particle flow are related to
$u^i$.
A point we wish to emphasize is that the term $wu^0u^i$ represents
the energy flux in the absence of viscosity.
The enthalpy density
depends on
 the material (particles
and antiparticles) in the  respective volume.

For the case of nonzero dissipation, an additional term $\tau ^{i0}$
containing
viscosity coefficients (see [5]) is added to $T^{i0}$ in (\ref {T}).
The
new tensor still satisfies (\ref{flo}). However, the particle
conservation
relationship (\ref{nflow}) is
replaced by:
\begin{equation}
\frac{\partial (nu^i + \nu ^i)}{\partial x^i}=0
\end{equation}
where $\nu ^i$ is proportional to the coefficient of thermal
conduction [5].
In the
relativistic dissipative case it is convenient to associate $u^i$
with
the velocity of energy flow, and to note that an
energy flux necessarily involves a mass flux [5].

 Next we examine the case of zero baryon number, i.e. the case of
zero
chemical potential and $n=0$. In this case there is no
net particle number and the thermal conductivity
is not defined, because there is no
net baryon density with respect to which energy can be conducted.
 The energy flow, however, need not be zero. The energy flow is
explicitly contained in (\ref{flo}).
In the low velocity limit, i.e. to first order in $v$ and in the
absence of viscosity, equation (\ref{flo}) gives:
\begin{equation}
\label{enf}
\partial _t e = -\nabla \cdot (w \vec{v})
\end{equation}
and for $i \not= 0$
\begin{equation}
\label{vfl}
 \partial_t w \vec{v} = - \nabla p.
\end{equation}
Here, the velocity of light has been set equal to $1$.
We wish to emphasize that
the energy flow, $w\vec{v}$, given by (\ref{enf}), does not vanish,
even
 though there
is no heat conduction.

 In the presence of viscosity, terms of second
order in $u$ appear in the energy equation (\ref{enf}), while a term
linear in  $u$
appears in (\ref{vfl}), the momentum equation. This means that the
viscosity
 terms are relatively
unimportant in the energy transport when $u$ is small. The
momentum equation, however, indicates that viscosity influences the
time evolution of $u$. Thus viscosity can serve to disrupt the energy
flow
 and generate
entropy but cannot be a
driving mechanism for energy removal.

 The authors of Ref. [1] also list equation (\ref{enf}) with the
high order viscosity terms dropped. This  explicitly
shows that energy can change in time in the absence of viscosity.
In Ref. [1], however, a different
equation
 with the dominant
term $\propto wu^i$ missing, and with the second order viscosity term
present, is also given and used to describe the rate of energy
change,
\begin{equation}
\label{vi}
\Delta w \frac{dR}{dt} = (\frac{4}{3} \eta + \zeta)
u_R\frac{du_R}{dr} .
\end{equation}
The terms $\eta$ and $\zeta$ are the shear and bulk viscosity
coefficients respectively,
 and the left hand side represents the energy
flux density (energy per unit area per time).

Equations (\ref{vi}), and (\ref{enf}) are inconsistent.
Equation (\ref{vi})suggests that the energy flow is provided
solely by viscous effects. It further suggest that,
in the
absence of such effects,  the hadron bubble could not grow
because the flow of the latent heat generated by the phase transition
is forbidden.
 The  fact that a $\vec v$ is present, however, means
there is energy flow  $\propto wv^i$, regardless of viscous effects,
and the fact that heat conduction can not be defined.
Physically one may
suggest  that the energy is being carried
away by quarks and antiquarks i.e. by convection to list a possible
mechanism.

\section{Solution of Relativistic Fluid Equations For Nonviscous
Case}

We now consider solving explicitly for the rate $\kappa$
in the absence of both
viscosity and baryon number. This exercise, explicitly shows
that $\kappa$
does not vanish. In Section 5, we further suggest
that the addition of viscous effects
leads to  small
corrections in $\kappa$.

In the nonviscous case and in the limit of small velocities
the relativistic energy and momentum differential equations become
[5] (\ref{enf}) and (\ref{vfl}).
  These equations have the same form whether the baryon
number is zero or not.

The method of solution we use here was suggested by C. Goebel.
The differential equations are solved in the quark vapor region only
and are then related to the interior hadron material through
boundary conditions.
 One begins with the
linearized relativistic equations in the quark region for a
velocity
field with spherical symmetry:

\begin{eqnarray}
\label{ee}
\partial_t \tilde {e} + \frac{1}{r^2}w_v \partial_r (r^2 v)&=&0 \cr
w_v \partial_t v + \partial_r \tilde {p} &=& 0.
\end{eqnarray}
The boundary conditions or Kotchine conditions, across the interface
are found by equating $T^{0i}$ in the two regions in a frame in which
the surface of discontinuity is at rest
and also $T^{ii}$ in
the two regions with a surface tension term added. The results are:
\begin{eqnarray}
\label{tf}
p_h - p_v &=& \frac{2 \sigma}{R} \cr
w_v(v_v-\dot{R})&=&w_h(v_h-\dot{R}).
\end{eqnarray}
Use has been made of the Galilean transformation as $v<<1$ and only
terms linear in $v$ retained. The $v's$ above represent velocities
measured relative to the bubble center.
Here the subscript $v$ refers to the quark vapor and $h$ to the
hadron
vapor. $\dot R$ is the velocity of the bubble surface. For simplicity
the velocity
inside the hadron bubble, $v_v$ is taken to be zero for the following
solution. This condition is not an essential one (see Appendix A).
A third boundary condition is needed to give information about the
combustion rate. We postulate it as:
\begin{equation}
w_v(v_v-\dot{R}) =C\Delta (\delta T_h-\delta T_v).
\end{equation}
Here $C$ is a constant related to the efficiency of energy flow
across
the quark-hadron interface. For illustration purposes we take:
\begin{equation}
\delta T_h=\delta T_v
\end{equation}
which is equivalent to $C \rightarrow \infty$. In a future work we
examine the effect of finite $C$.
 Since
 both $e$ and $p$ are functions of the temperature,
$T$,alone, the variations in $e$ and $p$, i.e., $\tilde e$ and
$\tilde p$, satisfy
\begin{equation}
\label{pstuff}
 \tilde {p} = c^2 \tilde {e}
\end{equation}
 where
\begin{equation}
 c^2 = p'_v/e'_v,
\end{equation}
 the prime denoting the $T$ derivative.
It can be shown that $c$ is the velocity of sound in the vapor.
 Next one can considers a variation of pressure
away from the stationary
solution, $\tilde p$. Then:
\begin{equation}
\tilde {p_v} - \tilde {p_h} = \frac{2 \sigma \tilde {R}}{R_o^2}.
\end{equation}
Here $R_o$ is the critical size and $\tilde {R}$ the variation from
this
size.
The quantities are all evaluated at $R_o$ in (\ref{tf}).

 Given the following forms for vapor and hadron pressures:
\begin{eqnarray}
\label{nos}
p_v &=& a_v T^4 -B \cr \cr
a_v&=&\frac{37 \pi^2}{90} \cr \cr
B&=& 235^4 Mev^4 \cr
p_h&=& a_h T^4 \cr \cr
 a_h&=&\frac{3 \pi^2}{90}.
\end{eqnarray}
one finds:
\begin{equation}
\label{pag}
\tilde {p_v} =- \frac{a_v}{a_v-a_h} \frac{2 \sigma}{R_o^2} \tilde{R}
\end{equation}
if one takes  $\delta T_v =\delta T_h$. This particular
choice provides for changes in pressure in both the hadron
bubble and quark vapor regions as the bubble grows.
 Differentiating
(\ref{pstuff}) with respect to time and using (\ref{pag}) one obtains
a relation which
holds at $R_o$
:
\begin{equation}
\label{flow}
\dot {\tilde {e}} = y z
\end{equation}
where
\begin{equation}
\label{wawa}
y = \frac{2 \sigma}{c^2 R_o^2}  \frac{w_v}{(\Delta w)^2}
\end{equation}
and
\begin{equation}
z = w_v v
\end{equation}
if $v_h$ is taken to zero as a first approximation (see Appendix A).
In the vapor region (\ref{flow}) becomes a wave equation upon
differentiation with respect to time and substitution of the second
relation of (\ref{ee}) differentiated with respect to $r$ :
\begin{equation}
\label{hit}
\partial _t ^2 \tilde{e}=c^2 \nabla ^2 \tilde{e}.
\end{equation}
 Explicitly using:
\begin{equation}
\label{goeb}
\tilde {e} = \frac{f' (r-ct)}{r}
\end{equation}
leads to:
\begin{equation}
z = \frac{c}{r^2} (rf'-f).
\end{equation}
At this stage $f$ is an arbitrary function and $f'$ is its derivative
with respect to $r-ct$. Using (\ref{flow}) one obtains:
\begin{equation}
f'' + yf' - \frac{y}{R_o}f =0.
\end{equation}
Solving with the form $f(x)={\rm exp}(\mu x)$ one obtains:
\begin{equation}
\label{hoha}
\mu = \frac{-1}{2} (y \pm \sqrt (y^2  +\frac{4y}{R_o}) .
\end{equation}

The upper sign is associated with the growing mode. Finally,
the rate $\kappa$ can be associated with $-c\mu$, so that
\begin{equation}
\label{res1a}
\kappa = c\frac{1}{2}(y + \sqrt (y^2 +\frac{4y}{R_o}).
\end{equation}

In the limit of small $y$ one obtains:
\begin{equation}
\label{res1}
\kappa = \sqrt \frac{2 \sigma w_v}{R_o^3 (\Delta w)^2}.
\end{equation}

For the purpose of comparison, we next review the solution provided
in Ref. [3] for a nonrelativistic system with nonzero chemical
potential.  That work provides a $\kappa$ for a drop (e.g. water)
growing in a nonviscous vapor with perfect heat conduction. (
Note: There is an error in the second part of [3] which involves
thermal conduction. The error is corrected in [7] and is unrelated
to the first evaluation of $\kappa$ in [3] which ignores changes
in temperature and hence also the specific thermal conduction
terms.)
 The relativistic equations (\ref{enf},\ref{vfl}) solved above are of
the
same
 form as the
nonrelativistic mass conservation and momentum equations used in the
first
 part of [3].
Thus one can compare the methods of solution and the results for
 the nonrelativistic treatment  of [3]  with
the relativistic case.

The nonrelativistic equations of [3],  expanded about a stationary
solution,
 are:
\begin{eqnarray}
\label{twefo}
\partial_t \nu &=& -\nabla \cdot (\bar n \vec {u}) \cr
\bar{n} \partial _t \vec{u}&=& -\nabla \tilde p.
\end{eqnarray}
Here $\bar {n}$ is the stationary density and $\nu$ and $\vec {u}$
the
changes in density and velocity from the stationary results. It is
further assumed that the time dependence of all quantities is
${\rm exp}(\kappa t)$. The term $\nabla \tilde p$ is obtained in [3]
from
the following expression for the free energy:
\begin{eqnarray}
F&=&F_K + F_I \cr
F_K&=& \frac{1}{2} \int d^3r m v^2 \cr
F_I&=& \int d^3r [\frac{1}{2} K(\nabla n)^2 + f(n)]
\end{eqnarray}
together with:
\begin{equation}
\label{ccc}
\frac{1}{n} p=\nabla \frac{\delta F_I}{\delta n}
\end{equation}
where $n$ is evaluated at $\bar{n} + \nu$. Equation (\ref{ccc}) is
consistent with the thermodynamical relation:
\begin{equation}
p=\mu N-f
\end{equation}
where
\begin{equation}
\mu=\frac{\partial F}{\partial N}
\end{equation}
holding $T$ and $V$ constant. Using
\begin{equation}
F=f(\frac{N}{V})V
\end{equation}
where $N$ is constant one can obtain $p$ from
\begin{equation}
p=-\frac{\partial F}{\partial V}
\end{equation}
holding $N$ and $T$ constant. Thus
\begin{equation}
p=n\frac{\partial f}{\partial n} -f.
\end{equation} One may then obtain:
\begin{equation}
\label{bana}
\partial_t \vec {u} = - \frac{1}{m} \nabla (-K\nabla^2 +
\frac{\partial^2 f}
{\partial \bar {n}^2}) \nu.
\end{equation}
It is interesting to note that although the expressions for the
pressure in terms of $f$ are different for the zero-baryon number
relativistic and nonzero baryon number nonrelativistic cases the
change in pressure
similar. For the two cases one has:
\begin{eqnarray}
\tilde{p}&=&c_1^2 \tilde{e}\cr
\tilde{p}&=&c_2^2 \tilde{n}
\end{eqnarray}
respectively. In both cases $c_i^2$ denotes the speed of sound in the
vapor. For the nonrelativisitic case:
\begin{eqnarray}
c_2^2&=&\frac{\partial p}{\partial n} \cr
&=& n\frac{\partial ^2 f}{\partial n^2}.
\end{eqnarray}

The first equation of (\ref{twefo}) is of the same form as the first
equation of (\ref {ee}) but the two have different origins. The
latter
is the result of energy conservation which follows from (\ref{flo})
involving
the energy-stress tensor. The former is a result of particle
conservation. There exists in relativistic fluid mechanics for
systems
with nonzero chemical potentials an equation
similar to the first of (\ref{twefo}) which describes the
conservation
of a net number of either particles or antiparticles. Such an
equation
only holds for nonzero baryon number and is not used in this paper
which deals with the zero baryon number case only. As a result it is
better not to look for a one-to-one correspondence between the
variables of the relativisitic and nonrelativistic equations even
though (\ref{twefo}) and (\ref{ee}) are of similar form with $n$
replacing $w$.

 One can proceed to solve the wave equation resulting from a
manipulation of (\ref{twefo}) in all space instead of solving just in
the exterior region and using the Kotchine boundary conditions as
done in
the method of section 3. The wave equation is straightforward to
solve
in the exterior and interior regions of the bubble.
 In
the interface region, however, the solution is more difficult as
$\partial^2 f/ \partial \bar {n}^2$ varies and Langer and Turski [3]
solve it approximately using the assumption of small $\kappa$.
Finally
one may integrate $\nu$ over all space and set this value equal to
zero because the total amount of material is conserved. The integral
in the interior contributes little and is ignored in [3]. The result
of
[3] then gives:
\begin{equation}
\kappa = \sqrt{ \frac{2 \sigma n_v}{m R_o^3 (\Delta n)^2}}.
\end{equation}

Here $n_v$ is the vapor density. It is interesting to note that
$\kappa$ vanishes with vanishing $n_v$. The factor $n_v/\Delta n$
enters the problem in the following way. In the interface region the
solution of $r\nu$ taken from [3] is equal to:
\begin{equation}
(\frac{\partial ^2 f}{\partial \bar n ^2})^{-1} = \frac{R^2 \Delta
n}{2\sigma} \frac{d\bar n}{dr}
\end{equation}
at $r$ near $R$.
Thus:
\begin{equation}
n_v \frac{\partial ^2 f}{\partial \bar n ^2} \nu= n_v \frac{2
\sigma}{R^2
\Delta n} (\frac{d \bar n}{dr}) ^{-1} \nu.
\end{equation}
The LHS is $\delta p_v$ which is equivalent to $\tilde{p_v}$ and the
factor $n_v/ \Delta n$ is
related to the expression for the pressure of the vapor in
equilibrium
with a drop in this case namely:
\begin{eqnarray}
p_v (total) &=& p_o + \delta p_v \cr
&=& p_o + \frac{2 \sigma n_v}{R_o \Delta n}
\end{eqnarray}
where $p_o$ is the pressure of a system with a large amount of liquid
in equilibrium with vapor. The term $\delta p_v$ describes the change
from this pressure due to the effects of the surface of the drop.
The procedure of solving in the interface region simply ensures that
the outside and inside pressures are related by:
\begin{equation}
\frac{2\sigma}{R}.
\end{equation}

One may now attempt to apply the above ideas to the relativistic
equations
(\ref{enf},\ref{vfl}). These lead to:
\begin{equation}
\partial _t ^2 \tilde{e} = - \nabla ^2 \tilde {p}.
\end{equation}
In the nonrelativistic and zero-baryon number relativistic
problem $\nabla \tilde{p}$ is equal to a constant multiplied by $\nu$
and $\tilde{e}$ respectively. Furthermore the same Kotchine equations
(\ref{tf}) with $w$'s replaced with $n$'s apply to the
nonrelativistic
problem. As a consequence the low $\kappa$ relativistic zero-baryon
result must be
the same as the nonrelativistic one with $n$'s replaced with $w$'s or
namely:
\begin{equation}
\kappa =\sqrt{( \frac{2 \sigma w_v}{R^3 (\Delta w)^2}}.
\end{equation}
This is identical to (\ref{res1}).

\section{Free Energy and Pressure Gradients}

  The work of Ref. [1] suggests that energy is removed during bubble
growth by viscous effects alone (\ref {vi}). This result is already
suggested by the authors' treatment in [1] of the relativistic
momentum
equation, i.e. the second of (\ref{ee}) with an additional term
linear
in viscosity coefficients to describe viscous effects. The result of
[1] that $\kappa$ is proportional to the viscosity coefficients
follows from the replacement of $\nabla p$ in (\ref {ee}) with a
particular expression involving the free energy of the bubble plus
vapor system. We examine the relationship between pressure and free
energy more closely to see how it leads to the conclusions of [1]
concerning
the role of viscosity.

 The free energy is taken to have the form \cite{hill}:
\begin{eqnarray}
\label{pu}
F&=& F_K + F_I \cr \cr
F_K &=& \frac{1}{2} \int d^3r w v^2 \cr \cr
F_I &=& \int d^3r [\frac{1}{2} K(\nabla e)^2 + f(e)].
\end{eqnarray}
Here $K$ is a constant related to $\sigma$, the surface tension. The
free energy is of the same form as that used in nonrelativistic work
[3] but is taken as a function of $e$, the energy density, as opposed
to $n$,
the number density. In the baryon-free case there is no number
density.
In [1] (equations 17-25) the pressure gradient term is associated
with:
 \begin{eqnarray}
\label{difpres}
-'\nabla p' &=& \frac{\delta F_K}{\delta e(\vec {r}}) \nabla e \cr
\cr
 &=& -K(\nabla^2 e) \nabla e + \frac{\partial f}{\partial e} \nabla
e.
\end{eqnarray}
 It is important to note that the $' \nabla p '$ is not simply a
pressure but a combination of a pressure and a force term.
 Both of these terms are found in the Euler or Navier-Stokes
equations. Such an identification must be made for the following
reason. The pressure on the inside differs from that on the outside
so
there is necessarily a pressure gradient in the interface. If there
were no force term then the Euler equation would require a changing
fluid velocity in the equilibrium situation which is unphysical. The
force term is associated with the $K (\nabla e)^2$ term which is
related to the surface tension by:
\begin{equation}
4 \pi R^2 \sigma = K \int d^3r (\nabla e)^2.
\end{equation}
This is the term needed to balance the differing pressures.
Identical considerations hold in the nonrelativistic case, with $n$
taking
the place of $e$.

 In a region of $e(\vec {r})$
varying slowly equation(\ref{difpres}) suggests the identification:
\begin{equation}
\label{iden}
\frac{\partial f}{\partial e} \nabla e = \nabla f = -\nabla p
\end{equation}
and
\begin{equation}
p=-f.
\end{equation}
 The term proportional to $K$ has been dropped in this region. We
also
assume this form.

 Let us examine
the approach used in [1]. There one linearizes the relativistic
momentum
equation for small variations, $\tilde{e}$ for the energy and
$\vec{v}$ for
the velocity, about the stationary solutions of $\bar{e}$ and
$\vec{0}$ respectively for the hadron bubble in
equilibrium with the quark vapor. Specifically:
\begin{equation}
\label{gah}
-\nabla \tilde{p}=\frac{\partial f}{\partial \bar{e}} \nabla \tilde
{e} +\frac{\partial ^2 f}{\partial e^2} \tilde{e} \nabla \bar{e}.
\end{equation}
The assumption:
\begin{equation}
\label{yt}
\frac{\partial f}{\partial e}=0
\end{equation}
is used. This is a consequence of the minimisation of (\ref{pu}) with
respect to $e$ to obtain a profile for the stationary solution
$\bar{e}$. The stationary solution of [1] is described by the
solution
of:
\begin{eqnarray}
\frac{\partial F_I}{\partial e}&=&-K\nabla ^2 \bar{e} +
\frac{\partial
f}{\partial \bar{e}} \cr
&=&0
\end{eqnarray}
where a fourth order polynomial in $e$ is chosen for $f$. This
approach follows nonrelativistic work where $n$ is used in place of
$e$. In the nonrelativistic case, however, one constrains the number
of particles present and so does not use:
\begin{equation}
\frac{\partial f}{\partial n}=0.
\end{equation}
With the use of (\ref{yt}) equation (64) of
[1] :
\begin{equation}
\label{k}
\partial_t (\bar{w} \vec{v})=(\nabla \bar{e})[-K(\nabla ^2) +f'']
\nu+ \nabla [ (\zeta + \frac{4}{3} \eta) \nabla \cdot \vec{v}]
\end{equation}
is obtained.
 One next makes the association $\partial_t
\vec {v} = \kappa \vec {v}$.  Evaluating (\ref{k}) in the quark vapor
region
where $\bar e$ is constant causes the first term in (\ref{k}) to
vanish
and suggests  $\kappa \propto$ viscosity coefficients.

Following this approach the velocity would not change in time and
the bubble would not grow in the absence of viscosity. We evaluate
(\ref{gah}) differently. The term:
\begin{equation}
\frac{\partial f}{\partial \bar{e}}
\end{equation}
is not taken to be zero but rather $c^2$. This does not cause $\nabla
\tilde{p}$ to vanish and so bubble growth occurs with or without
viscosity.

\section{Equations With Viscosity Present}
  In the presence of viscosity a term $\tau^{ik}$ is added to the
tensor
$T^{ik}$ (\ref{T}) where:
\begin{eqnarray}
\label{tooo}
\tau^{ik}&=& -\eta(\frac{\partial u^i}{\partial x^k} + \frac{\partial
u^k}{\partial x^i}+ u^k u^l \frac{\partial u^i}{\partial x^l} + u^i
u^l \frac{\partial u^k}{\partial
x^l}) \cr
&-&(\zeta-\frac{2}{3}\eta)\frac{\partial u^l}{\partial x^l} (g^{ik}
+u^i u^k).
\end{eqnarray}
The energy and momentum fluid equations in
the presence of viscosity, keeping only first order terms in $v$ the
velocity, are:
\begin{eqnarray}
\label{trrr}
\partial_t e &=& -\nabla \cdot w \vec {v} + O(v^2) \cr
w \partial_t v &=& -\nabla p + \nabla[(\frac{4}{3}\eta +\zeta)\nabla
\cdot \vec{v}]
\end{eqnarray}

These are the same as those of [1].
 For small viscosity
coefficients and small $\kappa$ it appears that the viscosity term is
small compared to other terms as it
involves two gradients. Here we assume a non-vanishing pressure
gradient.
 From the equations with viscosity absent
one can see that gradients of $v$ are smaller than $e$ by a factor of
$\kappa$, and $\kappa$ is assumed to be small.
Next we consider the effect of viscosity on the boundary
conditions. In the absence of viscosity the boundary conditions
(\ref{tf}) can be expressed, in a frame moving with the interface, by
\begin{eqnarray}
\frac{x_i x_j}{r^2}T^{ij}_v &=& \frac{x_i x_j}{r^2}T^{ij}_h \cr \cr
\frac{x_i}{r} T^{0i}_v&=&\frac{x_i}{r}T^{0i}_h.
\end{eqnarray}
 In the presence of viscosity:
\begin{equation}
T^{ij}=pg^{ij}+\tau ^{ij}
\end{equation}
to first order in velocity where only first order velocity terms in
$\tau$ are retained.

 In a spherical problem one has
\begin{eqnarray}
\label{nbc}
T^{ij}\frac{x_jx_i}{r^2}&=& \cr
&&[p+(\frac{4}{3}\eta +\zeta)\nabla \cdot (v-\dot{R}) \cr
 &+&\frac{4\eta(v-\dot{R})}{r}]
\end{eqnarray}
Thus $p$ in the first of (\ref{tf}) is replaced with the RHS of
(\ref{nbc}) while the
 second boundary condition of (\ref{tf}) remains unchanged to first
order in velocity. Here $v$ is the velocity measured in a frame in
which the bubble center is at rest. A Galilean transformation has
been performed to relate this frame with that moving with the
interface because all velocities are small compared to that of light.

We first consider the effect of the viscosity-modified transport
equations.
 One can solve (\ref{trrr})
explicitly following the method of section 3, using (\ref{pstuff}h)
and replacing $-\nabla \cdot
\vec{v} $ in (\ref{trrr}) with $\partial_t e /w$. The equation in the
vapor
becomes:
\begin{equation}
\partial_t ^2 \tilde{e} = c^2 \nabla ^2 \tilde{e} + \nabla ^2
\frac{(\frac{4}{3} \eta + \zeta)}{w_v} \partial_t \tilde{e}.
\end{equation}
For $\frac{(\frac{4}{3}\eta+\zeta)}{w_v}$ constant in space one
obtains:
\begin{equation}
\label{elll}
\tilde{e}=\frac{A}{r} {\rm exp}(-qr+ \kappa t)
\end{equation}
where $A$ is a constant and $\kappa$ and $q$ are related by:
\begin{equation}
\label{pp}
\kappa ^2 = c^2 q^2 + \frac{(\frac{4}{3} \eta + \zeta)}{w} \kappa
q^2.
\end{equation}
Thus in the vapor region one obtains the same result as (\ref{hit})
but
with $c^2$, the speed of sound, replaced with:
\begin{equation}
c^2(1+  \frac{(\frac{4}{3} \eta + \zeta)}{w} \frac{\kappa}{c^2}) .
\end{equation}
Due to the presence of a $\kappa$ factor the second term on the RHS
of
(\ref{pp})is
small compared to the first. This suggests
that viscosity is a small effect.
Solving the first of (\ref{ee}) which is the same for both the
nonviscous and viscous situations
an expression for $v$:
\begin{equation}
\label{velo}
v=\frac{\kappa}{wq^2} A(\frac{q}{r} +\frac{1}{r^2}) {\rm
exp}(-qr+\kappa t)
\end{equation}
is found.

 To see the explicit effect of viscosity on $\kappa$ a relationship
incorporating the boundary conditions is needed. Eq. (\ref{pp})
simply
provides a relationship between two unknowns $q$ and $\kappa$;
an additional equation is needed relating the two.
 To obtain such an expression
  (\ref{ee})
 and (\ref{nbc}) are combined in the same manner as (\ref{ee}) and
(\ref{tf}) in secton 3 to obtain an expression analagous to
(\ref{pag}). The
addition
 of viscosity modifies (\ref{pag}). To see how this modification
arises we consider the derivation of (\ref{pag}) in section 3.
To obtain the result (\ref{pag}) one solves for $\delta T_h=\delta
T_v
\equiv \delta T$ in the first equation of (\ref{tf}) by setting:
\begin{equation}
\label{pst}
\tilde{p_v} \propto a_v \delta T_v
\end{equation}
and taking a similar expression for $\tilde{p_h}$. One then
substitutes the
 value
for $\delta T$ in (\ref{pst}) to obtain $\tilde{p_v}$. In the
presence
of viscosity a similar method is employed but now $u_h,u_v,\dot{R}$
in
 (\ref{nbc}) must be replaced with
expressions involving $\delta T$. This allows one to solve for
$\delta
T$ and thus $\tilde{p_v}$.
Terms with $\nabla \cdot v$ can be
replaced with terms involving $\kappa \tilde{e}$ using the first of
(\ref{trrr}). Terms involving $\nabla \cdot \dot{R}$ are
proportional to $\kappa \tilde{e_v}$
  because the second of (\ref{tf}) relates $\dot{R}$ to
$v_v$. Terms involving $v_h$ are dropped to be consistent with
section
3 (see Appendix A)
. Finally one must consider terms with $\dot{R}$
alone. Given (\ref{elll}) and (\ref{velo}) it is possible to relate
$v_v$ and
$\tilde {e_v}$ at the interface to obtain:
\begin{equation}
v_v=\frac{\kappa}{q^2 w_v} (q+\frac{1}{R})\tilde{e_v}.
\end{equation}

 The use of (\ref{nbc}) essentially
 leads to the  replacement of the factor:
\begin{equation}
\frac{a_v}{a_v-a_h}=\frac{w_v}{\Delta w}
\end{equation}
appearing in (\ref{pag}) with the factor
\begin{eqnarray}
x&\equiv&w_v/[w_h-w_v(1+\frac{3f_v\kappa}{\Delta
w}-\frac{3f_q\kappa}{\Delta w}+\frac{3f_q \kappa }{w_q} \cr
&&+\frac{12c^2N^2}{
R\Delta w w_v \kappa}(\eta _h w_v -\eta _v w_h)(\frac{\kappa}{cN}
+\frac{1}{R_o}))].
\end{eqnarray}
 Here
\begin{eqnarray}
f_h&=&\frac{4}{3}\eta _h +\zeta _h \cr
f_v&=&\frac{4}{3}\eta _v +\zeta _v \cr
N&=&\sqrt {(1 + \frac{(\frac{4}{3} \eta + \zeta)}{wc^2} \kappa)}.
\end{eqnarray}

Next Eq. (\ref{pstuff}) is differentiated with respect to time to
obtain an expression for $\dot{\tilde{p_v}}$.
 The term $\dot{\tilde p}$ (\ref{pstuff}) is replaced with an
expression involving $\dot {\tilde R}$ which is in turn replaced with
an expression in the velocity $v$. The term $\dot {\tilde e}$ is
written in terms of the gradient of the velocity. One then obtains:
\begin{equation}
\label{catt}
 -\frac{1}{r^2} w_v \partial _r r^2 v = Yw_v v
\end{equation}
where:
\begin{equation}
Y=yx\frac{\Delta w}{w_v},
\end{equation}
$y$ being given in (\ref{wawa}).
 The velocity in the hadron material
is again neglected.

Substituting into (\ref{catt}) yields:
\begin{equation}
\label{hat}
 Y(q+\frac{1}{R_o}) = q^2
\end{equation}
which may be compared with (\ref{res1a}), the result in the absence
of viscosity.
Eq. (\ref{hat}) must be solved with $q$ expressed in terms of
$\kappa$  namely:
\begin{equation}
q=\frac{\kappa}{cN}.
\end{equation}
The modification due to viscosity resulting from the change in the
transport equations (\ref{trrr}) is present in $q$ while that due to
the change in the boundary conditions (\ref{nbc}) is contained in the
replacement of $y$ with
$Y$. In general (\ref{hat}) must be solved numerically for $\kappa$
as
it is of higher than third order in $\kappa$. An approximate solution
is obtained if
  the $q$ term can be neglected
compared to $\frac{1}{R_o}$ in (\ref{hat}):
\begin{equation}
\label{shq}
\kappa = cN\sqrt{\frac{Y}{R_o}}
\end{equation}
with $Y$ and $N$ being evaluated with $\kappa$ replaced with $\kappa
_o$, the result in the absence of viscosity
(\ref{hoha}). The solution for $\kappa$ in the absence of viscosity
is
of the same form as (\ref{shq}) except $N$ is replaced with $1$ and
$Y$ with $y$.

\section{Conclusion}

We have examined the growth of a hadron bubble in quark vapor and
suggest
 that the physical process is different from that presented in
[1]. Specifically we suggest that the formalism of relativistic
hydrodynamics permits energy flow in the absence of heat conduction
and viscosity. We have solved the relativistic fluid equations for a
system with
zero baryon number (no thermal conduction) and no viscosity and
have obtained a nonzero value for $\kappa$. We have compared the
relativistic fluid equations and resulting $\kappa$ with the
nonrelativistic equations and $\kappa$ presented in [3]. The method
used in [3] is approximate in that it only holds in the limit of
small
$\kappa$ and differs from that presented in this
paper which provides an exact result for $\kappa$. Finally we have
solved the relativistic fluid equations for zero baryon number in the
presence of viscosity. We suggest that viscous effects introduce only
a small correction in the nonviscous value of $\kappa$ and provide an
expression for the correction.

In this paper we have used a specific set of boundary conditions
linking the quark and hadron regions. Investigation of different sets
of boundary conditions and their effect on $\kappa$ is in progress.
In addition we consider the calculation of $\kappa$ for a medium with
nonzero baryon number and thus heat conduction.

\acknowledgements
We are grateful to C. Goebel for discussions and suggestions.
\appendix
\section{A}
   A solution for $\kappa$ has been obtained under the assumption
that $\delta T_h = \delta T_q$ at the surface of the bubble
and that the velocity in the hadron
material is zero. One may solve for $\kappa$ without the latter
restriction to find a more general result which in the limit of small
$\kappa$
reduces to the zero velocity solution. For small
$\kappa$ the velocity on the inside is much less than that on the
outside and so can be ignored. One may
note that the authors of [3], who consider the problem of
nonrelativistic drop forming in a vapor, also disregard the velocity
inside
the drop in the limit of small $\kappa$.

To solve the general problem  with velocity retained in both the
bubble and vapor regions the hydrodynamical equations
must be solved in both regions and use made of the
full boundary condition:
\begin{equation}
w_q v_q -w_q \dot{R}= w_h v_h - w_h \dot{R}.
\end{equation}
The solutions in the inner region are:
\begin{eqnarray}
\tilde{e_h}&=&C_h \frac{\rm{sinh} (qr)}{\rm{sinh} (qR_o)}
\frac{e^{\kappa t}}{r} \cr \cr
v_h&=&\frac{C_h}{\rm{sinh} (qR_o)} \frac{\kappa}{w_h q^2}
(-q\rm{cosh}(qr)
+\frac{\rm{sinh}(qr)}{r}) \frac{e^{\kappa t}}{r}.
\end{eqnarray}
In addition there is the relation:
\begin{equation}
\label{lave}
\kappa = q c_h
\end{equation}
where $c_h$ is the speed of sound in the hadron material and equals
that of the quark medium. The solution in the quark region is of the
form:
\begin{equation}
\tilde{e_q}=C_q \frac{\rm{exp}[\kappa t - q(r-R_o)]}{r}.
\end{equation}
Thus one may obtain the relationship:
\begin{eqnarray}
\frac{\tilde{e_q}}{\tilde{e_h}}&=&\frac{a_q \delta T_q}{a_h \delta
T_h} \cr \cr
&=& \frac{C_q}{C_h}
\end{eqnarray}
at the interface surface.
Finally one may solve for $\kappa$ by evaluation of:
\begin{eqnarray}
\dot {\tilde p_q} &=& c^2 \kappa \tilde{e_q} \cr
&=&\frac{2\sigma}{R^2 \Delta w} (w_q v_q -w_h v_h)
\end{eqnarray}
with (\ref{lave}). All quantities are evaluated at the surface of the
bubble. The resulting expression for $\kappa$ is complicated due to
the presence of the terms cosh$qR$ and sinh$qR$. One may, however,
consider the limit of small $q$ which is equivalent to small
$\kappa$. In that case:
\begin{equation}
\label{uh}
v_h \propto -\kappa R.
\end{equation}
Here the cosh and sinh terms have been expanded to second and third
powers of $q$ respectively.
It is possible to see that $v_h$ is much smaller than the expression
for
$v_q$ in the case of small $q$ and so $v_h=0$ as an
approximation. One can proceed to solve for $\kappa$ with the
next order approximation using (\ref{uh}) instead of $v_h=0$. Then
one
must solve the quadratic:
\begin{equation}
c^2 q^2=\frac{2\sigma w_q}{R^2 (\Delta w)^2} (q+\frac{1}{R}
+\frac{a_h
R q^2}{a_q 3})
\end{equation}
to obtain:
\begin{equation}
q=\frac{1}{2c^2(-y\frac{a_h R}{a_q
3})}[yc^2+\sqrt{c^4y^2-\frac{4yc^4}{R}(1-y\frac{a_hR}{3a_q}}].
\end{equation}
Here $y$ is given by (\ref{wawa}).
In the limit $y<<\frac{4}{R}$ one obtains the results of section 3.
\section{B}
A method to solve for $\kappa$ using differential equations in a
metastable and stable phase region together with Kotchine conditions
relating the two phase at the bubble surface is presented in section
3. This method holds $\sigma$, the surface tension, fixed as the
bubble
grows. An alternative method, which yields identical results, is
considered in this appendix. It is based on a modification of Langer
and Turski's approach to the
nonrelativistic problem. In the Langer and Turski approach, as
outlined in section 3, one solves a
wave equation for $\tilde{n}$ in the vapor and interior drop regions
and obtains
solutions for $\tilde{n}$ and $v$ in these two
regions. These solutions are functions of $\kappa$. Next an
approximate solution for $\tilde{n}$ is obtained in
the interface region assuming small values of $\kappa$.
Finally the relation:
\begin{equation}
\label{dencon}
\int d\vec{r} \tilde{n}(r)=0,
\end{equation}
namely particle conservation is used to obtain a value for
$\kappa$. In [3] only low order terms in $\kappa$ are retained after
integration. If one, however, evaluates the expression exactly and
considers the manipulations in the interface to also be exact
 one
obtains the same results as in section 3.
Specifically, in the nonrelativistic problem one has:
\begin{equation}
\partial _t \tilde{n}=-\nabla \cdot nv.
\end{equation}
 The condition
(\ref{dencon}) is equivalent to:
\begin{equation}
\label{time}
\int d\vec{r} \partial _t \tilde{n} =0
\end{equation}
for the case of interest namely:
\begin{equation}
\partial _t \tilde{n}=\kappa \tilde{n}.
\end{equation}
The volume of all space can now be separated into three
regions. Two of the regions are the inner drop and outer vapor
regions and
the third is the vicinity of the surface in which the density is
changing.
Using (\ref{time}) and Gauss' law one may make the identification:
\begin{equation}
\int \partial _t \tilde{n}= interface piece + n_{in} v_{in}
-n_{out}v_{out}.
\end{equation}
The surface piece is equivalent to:
\begin{equation}
\label{jp}
(\Delta n) \tilde{R}.
\end{equation}
This describes the change in the interface as equivalent to a drop
with a sharp surface growing by the amount $\tilde{R}$. The inner and
outer areas defining the interface are taken as equal.
 An expression
for $\tilde{R}$ can be obtained from:
\begin{equation}
\label{jq}
\tilde{p}_{vapor}=-\frac{2\sigma n_v}{R^2(\Delta n)^2} \tilde{R}
\end{equation}
and
\begin{equation}
\tilde{p_v}=c^2\tilde{n}.
\end{equation}
Here the Kotchine condition relating pressures on the inside and
outside regions has been used.
 Langer and
Turski in [3] obtain a similar result by approximately solving the
transport equation:
\begin{equation}
m\kappa ^2 \tilde{n}=\nabla \cdot n \nabla(-K\nabla ^2
+\frac{\partial
^2f}{\partial n^2})\tilde{n}
\end{equation}
in the region in which the density is changing. The result of [3]
keeps only terms to first order in $\kappa$ but is equivalent to
(\ref{jp}) and (\ref{jq}).
Thus one finally obtains:
\begin{equation}
0=n_{in}v_{in}-n_{out}v_{out} +\frac{c^2\tilde{n}R^2\Delta n}{2\sigma
n_v}.
\end{equation}
which is identical to the result of section 3 obtained using a method
which utilised the Kotchine expressions and $\tilde{n}$ and $v$
solutions together with:
\begin{equation}
\tilde{p}=c^2 \tilde{n}
\end{equation}
Identical arguments hold for the relativistic problem.
 For the relativistic problem to first order in the energy flow
velocity over the speed of light one has:
\begin{equation}
\partial _t \tilde{e}=-\nabla \cdot wv.
\end{equation}
This equation holds with or without viscosity for both baryon and
baryon-free materials. Separating the volume again into three regions
and using Gauss' law leads to the energy conservation condition:
\begin{equation}
w_hv_h-w_qv_q=\Delta w \dot{R}.
\end{equation}

{}
\end{document}